\documentclass{camera}
\usepackage{graphicx}

\begin{document}
\title{Tracing a phase transition with 
fluctuations of the largest fragment size: 
Statistical multifragmentation models and 
          the ALADIN S254 data}

\author{T. Pietrzak$^{1}$, P. Adrich$^{2}$, T. Aumann$^{2}$, C. O. Bacri$^{3}$,\\ T. Barczyk$^{1}$,
R. Bassini$^{4}$, S. Bianchin$^{2}$, C. Boiano$^{4}$,\\ A. S. Botvina$^{2,5}$, A. Boudard$^{6}$,
J. Brzychczyk$^{1}$, A. Chbihi$^{7}$,\\ J. Cibor$^{8}$, B. Czech$^{8}$, M. De Napoli$^{9}$,
J.-\'{E}. Ducret$^{6}$, H. Emling$^{2}$, J. D. Frankland$^{7}$, M. Hellstr\"{o}m$^{2}$, D. Henzlova$^{2}$,
G. Imm\`{e}$^{9}$,\\ I. Iori$^{4,*}$, H. Johansson$^{2}$, K. Kezzar$^{2}$, A. Lafriakh$^{6}$,\\ A. Le F\`{e}vre$^{2}$,
E. Le Gentil$^{6}$, Y. Leifels$^{2}$, J. L\"{u}hning$^{2}$,\\ J. \L{}ukasik$^{8}$, W. G. Lynch$^{10}$,
U. Lynen$^{2}$, Z. Majka$^{1}$, M. Mocko$^{10}$, W. F. J. M\"{u}ller$^{2}$, A. Mykulyak$^{11}$, H. Orth$^{2}$,
A. N. Otte$^{2}$,\\ R. Palit$^{2}$, P. Paw\l{}owski$^{8}$, A. Pullia$^{4}$, G. Raciti$^{9,*}$,
E. Rapisarda$^{9}$, H. Sann$^{2,*}$, C. Schwarz$^{2}$, C. Sfienti$^{2}$, H. Simon$^{2}$,\\ K. S\"{u}mmerer$^{2}$,
W. Trautmann$^{2}$, M. B. Tsang$^{10}$, G. Verde$^{10}$,\\ C. Volant$^{6}$, M. Wallace$^{10}$, H. Weick$^{2}$,
J. Wiechula$^{2}$,\\ A. Wieloch$^{1}$ \and B. Zwiegli\'{n}ski$^{11}$}


\organization{
$^{1}$ Institute of Physics, Jagiellonian University, 30-059 Krak\'{o}w, Poland\\
$^{2}$ GSI Darmstadt, D-64291 Darmstadt, Germany\\
$^{3}$ IPN Orsay, IN2P3-CNRS et Universite, F-91406 Orsay, France\\
$^{4}$ Istituto di Scienze Fisiche, Universit\`{a} and INFN, I-20133 Milano, Italy\\
$^{5}$ Institute for Nuclear Research, 117312 Moscow, Russia\\
$^{6}$ DAPNIA/SPhN, CEA/Saclay, F-91191 Gif-sur-Yvette, France\\
$^{7}$ GANIL, CEA et IN2P3-CNRS, F-14076 Caen, France\\
$^{8}$ Institute of Nuclear Physics PAN, 31-342 Krak\'{o}w, Poland\\
$^{9}$ Dipartimento di Fisica and INFN-LNS, I-95123 Catania, Italy\\
$^{10}$ Dept. of Physics and NSCL, MSU, East Lansing, MI 48824, USA\\
$^{11}$ A. So\l{}tan Institute for Nuclear Studies, Pl-00681 Warsaw, Poland\\
$^{*}$ deceased}
\maketitle

\begin{abstract}

A phase transition signature associated with
cumulants of the largest fragment size distribution
has been identified in statistical multifragmentation models
and examined in analysis of the ALADIN S254 data on fragmentation
of neutron-poor and neutron-rich projectiles.
Characteristics of the transition point indicated by this signature
are weakly dependent on the A/Z ratio of the fragmenting spectator source.
In particular, chemical freeze-out temperatures are estimated within
the range 5.9 to 6.5 MeV. The experimental results are well reproduced
by the SMM model.
\end{abstract}

In nuclear multifragmentation studies
a special attention has been paid to the largest fragment size
which is expected to play the role of the
order parameter, and thus provide valuable insight
into the phase behavior of investigated systems \cite{bot,ma,frank,pi,jb}.
In particular, percolation studies have shown that a cumulant analysis
focused on the skewness, $K_{3}$, and the kurtosis excess, $K_{4}$,
of the largest fragment size distribution is a valuable
method to reveal the presence of a phase transition
(critical behavior) in finite systems \cite{jb}. The percolation transition is
indicated by $K_{3}=0$ and a $K_{4}$ minimum
of about $-1$ for all system sizes. Events may be sorted
according to various measurable quantities that are
correlated with the control parameter.

In this work we show that such cumulant features are not restricted
to the percolation transition but also are observed at a liquid-gas
phase transition present in statistical multifragmentation models.
With this justification the cumulant analysis will be applied to 
the ALADIN S254 experimental data obtained with
radioactive beams to investigate the presence and isotopic dependence of
the transition signals.

It is instructive to look first at the predictions of the
thermodynamic model which is known to contain a first-order phase transition.
The thermodynamic model is a simplified version of the statistical
multifragmentation model (SMM), allowing to compute the partition function
and thus to obtain thermodynamic properties of the system \cite{the}.
Our calculations were performed for the canonical ensemble
of noninteracting one-component fragments at the constant
freeze-out volume three times larger than the normal nucleus volume.
Figure 1 shows the specific heat and the statistical measures of
the largest fragment size distribution for three systems
with $64$, $216$, and $1000$ nucleons.
The broadening of $C_{V}$ peaks and reduction in peak intensities,
as well as the decrease of the transition temperatures indicated by the $C_{V}$ maxima,
is observed with decreasing system size due to finite-size effects. 

\begin{figure}[ht]
\begin{center}
\includegraphics[width=9.0cm]{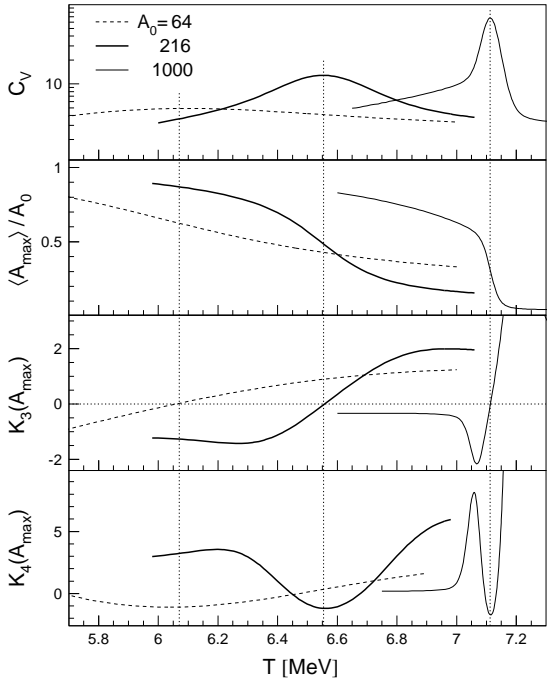}
\caption{Canonical thermodynamic model predictions. From top to bottom:
the specific heat, the mean, skewness and kurtosis of the largest fragment mass
distribution as a function of the freeze-out temperature for different
system masses.}
\label{fig01}
\end{center}
\end{figure}

At the transition point, the mean $A_{max}$
exhibits the fastest decrease, which tends with increasing system size to
a step discontinuity expected in the thermodynamic limit.
It is also clearly seen that the transition is associated with 
$K_{3}=0$ and minimum $K_{4}$, as in the case of the percolation
transition. Although these cumulant features are similar for both
transition types, the shapes of the transitional $A_{max}$ distributions
are distinctly different.
In the present model the distribution has a bimodal structure 
as expected for the order parameter at a first-order phase transition in the canonical
ensemble. In percolation the transition is continuous and the distribution
is single-peaked.

In the next step we have examined the cumulants of the largest fragment charge
($Z_{max}$) distributions within the SMM model \cite{smm}. Calculations were
performed with the standard SMM code which includes secondary decays.
The freeze-out density of one third of the normal nuclear density was assumed.
For a given system (A,Z), events were uniformly generated over a wide range
of the excitation energy, and then sorted according to the measurable quantity
$Z_{bound}$ used as a control parameter.The simulation results have shown that,
similarly as in the thermodynamic model, the cumulant signal is well observed
at a point where $Z_{max}$ rapidly decreases. The mean excitation energy
and the mean microcanonical temperature at this $Z_{bound}$ point correspond to
the flattest part of the caloric curve.
The temperature derived in this way, referred to as
the SMM breakup temperature, $T_{b}$, is shown in Fig. 2(a) as
a function of the system charge $Z$ for the $A/Z$ ratios of 2.17 and 2.49.
These ratios are relevant to the neutron-poor and neutron-rich systems
investigated in the S254 experiment. As one can see, the temperature is around
6 MeV and weakly depends on the system size and the $A/Z$ ratio.
For a comparison, Fig. 2(b) shows the Hartree-Fock limiting temperatures \cite{bes}.
In this case, regarding their absolute magnitude
which depends on the nuclear potential form used in the calculations \cite{baldo},
the mass/charge dependences are much stronger. The discrimination between
the two scenarios was one of the motivations for the S254 experiment.

\begin{figure}[ht]
\begin{center}
\includegraphics[width=10.0cm]{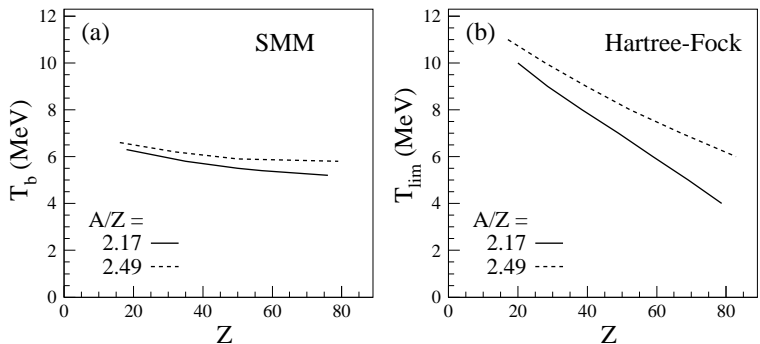}
\caption{SMM breakup temperatures (a) and limiting
temperatures from \cite{bes} (b).}
\label{fig02}
\end{center}
\end{figure}


The ALADIN experiment S254 was conducted at GSI Darmstadt.
Beams of $^{107}$Sn, $^{124}$Sn and $^{124}$La were used to investigate
isotopic effects in projectile-spectator fragmentation at 600 AMeV.
The secondary beams with neutron-poor $^{107}$Sn and $^{124}$La projectiles
contained also some fraction of neighbouring isotopes. The mean compositions of
the nominal $^{107}$Sn ($^{124}$La) beams were $\langle Z\rangle=49.7 \, (56.8)$ and
$\langle A/Z\rangle=2.16 \, (2.19)$, respectively \cite{luk}.

The most prominent result of the experiment is the observation that
the isotopic dependence of the projectile fragmentation is weak \cite{sfienti}.
The mean IMF multiplicity, the mean largest fragment charge, and the double-isotope
temperatures, $T_{HeLi}$ and $T_{BeLi}$, determined as a function of $Z_{bound}$,
are nearly invariant with respect to the projectile $A/Z$ ratio.
These can be seen in the top panels of Fig. 3 for the former
quantities. When comparing the Sn and La results, the resemblance
is better if $\langle Z_{max} \rangle$ and $Z_{bound}$ are normalized to
the projectile charge.
Figure 3 allows to examine the pattern of $Z_{max}$ fluctuations
by looking at the variance and the $K_{3}$, $K_{4}$ cumulants of the $Z_{max}$
distribution. Also in this case the results are similar for all the
projectiles.

\begin{figure}[ht]
\begin{center}
\includegraphics[width=10.0cm]{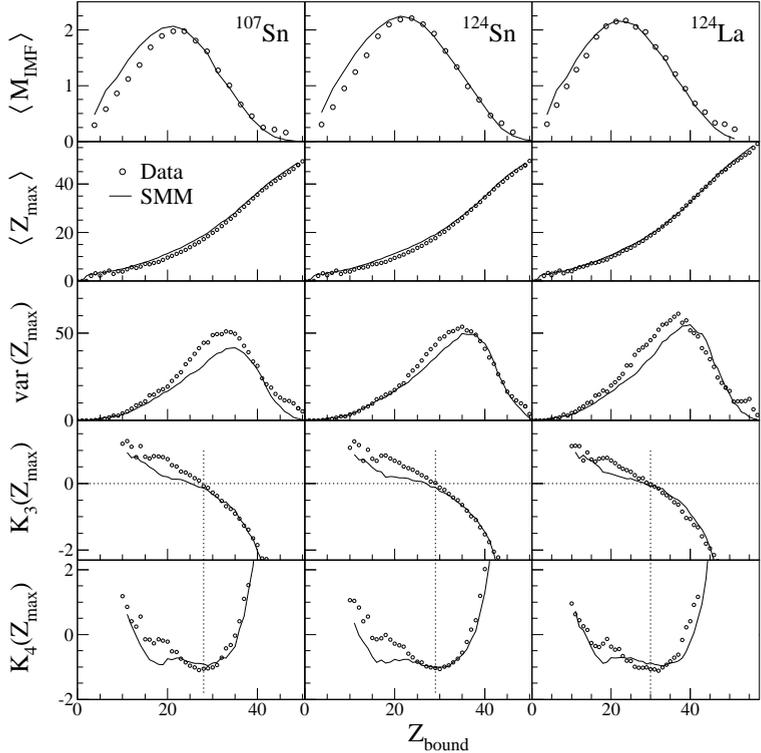}
\caption{Experimental data versus SMM predictions.
From top to bottom: the mean IMF multiplicity, the mean, variance,
skewness and kurtosis of the $Z_{max}$ probability distribution
as a function of $Z_{bound}$.}
\label{fig03}
\end{center}
\end{figure}

The signals characteristic of a phase transition
($K_{3}=0$ and minimum $K_{4}$) are observed
at $Z_{bound}=$ 28, 29 and 30 for the $^{107}$Sn, $^{124}$Sn and $^{124}$La
projectiles, respectively. It should be noted that the shapes of the
investigated $Z_{max}$ distributions are highly irregular since
the constraints imposed by fixed $Z_{bound}$ values are strong
in such small systems.
Therefore, the presence of some distribution features
({\it e.g.} bimodality) cannot be trustworthy concluded.
Despite that, the evolution of the cumulant values is quite smooth
so that the transition signals are clearly identified.
Given the signal locations on the $Z_{bound}$ axis, we could estimate parameters of
the fragmenting sources at the transition as shown in Table 1.
The mean charge of the fragmenting system, $Z_{0}$, was estimated
on the basis of percolation and SMM simulations as 
10-20\% larger than the $Z_{bound}$ value.
The values of the isotope temperature $T_{HeLi}$ were raised
by 15\% with a possible error of $\pm 5 \%$, according to
the expected range of corrections for secondary decays feeding \cite{traut}.
The obtained temperatures are in an approximate agreement with
the SMM predictions given in the last column.
The SMM breakup temperatures were read out from Fig. 2 with
the assumption that the $A/Z$ ratio of the
fragmenting system remains that of the projectile.

\begin{table}[ht]
\center
\small
\begin{tabular}{ccccc}
\hline
Projectile&$Z_{bound}$&$Z_{0}$&$T_{HeLi}$&$T_{b}$ - SMM \\
&&&(MeV)&(MeV) \\
\hline
$^{107}$Sn & 28&32.2${\pm}$2 & 5.9$\pm$0.3 & 5.87$\pm$0.05 \\
$^{124}$Sn & 29 & 33.5$\pm$2 & 6.5$\pm$0.3 & 6.18$\pm$0.05 \\
$^{124}$La & 30 & 34.5$\pm$2 & 6.2$\pm$0.3 & 5.82$\pm$0.05 \\
\hline
\end{tabular}
\caption{Freeze-out characteristics of the fragmenting systems
at the transition indicated by $Z_{max}$ fluctuations.}
\end{table}

For more detailed comparisons with the experimental
data we have performed SMM ensemble calculations following
the procedure described in \cite{botv}. The ensemble specifies
the probability distribution of the thermalized sources defined
by $A$, $Z$ and $E^{*}$.
Parameters of the ensemble were adopted from
earlier results for $^{197}$Au fragmentation \cite{botv},
and then individually adjusted to improve agreements with the experimental IMF multiplicity
distributions. The quality of the obtained agreements can be seen in the top
panel of Fig. 3. The next panels show the predicted characteristics of the $Z_{max}$
distributions. The overall agreement with the data can be concluded as satisfactory,
considering the fact that no optimization of the model parameters was performed.


In summary, simulations with the canonical thermodynamic model and the SMM model
corroborate the percolation suggestion that cumulants such as the skewness ($K_{3}$)
and the kurtosis excess ($K_{4}$) of the largest fragment size distribution
are valuable observables in searching for a phase transition in multifragmentation.
The location of a transition is well indicated by $K_{3} = 0$ and $K_{4}$ minimum,
even in very small systems. The measurable quantity $Z_{bound}$
may be used to sort events as a control parameter.
The cumulant analysis applied to the ALADIN S254 experimental data has shown that
the transition signals are observed at nearly the same $Z_{bound}$ values
of 28-30 for all the projectiles. The corresponding
freeze-out temperatures are in the vicinity of 6 MeV and
the temperature difference between the neutron-rich and neutron-poor sources
is within a few percent, in agreement with SMM predictions.
A satisfactory overall description of the experimental $Z_{max}$ distributions
can be obtained with SMM ensemble calculations.
The observed agreement with the statistical model indicate that in the
multifragmentation process the accessible phase-space is of dominant importance.

\smallskip

This work has been supported by
the European Community under Contract No. HPRI-CT-1999-00001
and the Polish Ministry
of Science and Higher Education grant N202 160 32/4308 (2007-2009).


\begin{thebibliography}{10}

\bibitem{bot}  {BOTET R. ET AL., {\it Phys. Rev. Lett.} {\bf 86} (2001) 3514.}

\bibitem{ma} {MA Y. G. ET AL., {\it Phys. Rev.} C {\bf 71} (2005) 054606.}

\bibitem{frank} {FRANKLAND J. D. ET AL., {\it Phys. Rev.} C {\bf 71} (2005) 034607.}

\bibitem{pi} {PICHON M. ET AL., {\it Nucl. Phys.} {\bf A779} (2006) 267.}

\bibitem{jb} {BRZYCHCZYK J., {\it Phys. Rev.} C {\bf 73} (2006) 024601.}

\bibitem{the} {DAS C. B., DAS GUPTA S., LYNCH W. G., MEKJIAN A. Z., TSANG M. B.,
{\it Phys. Rep.} {\bf 406} (2005) 1.}

\bibitem{smm} {BONDORF J. P., BOTVINA A. S., ILJINOV A. S., MISHUSTIN I. N.,
SNEPPEN K., {\it Phys. Rep.} {\bf 257} (1995) 133.}

\bibitem{bes} {BESPROVANY J. AND LEVIT S., {\it Phys. Lett.} B {\bf 217} (1989) 1.}

\bibitem{baldo} {BALDO M., FERREIRA L. S. AND NICOTRA O. E., {\it Phys. Rev.} C {\bf 69} (2004) 034321.}

\bibitem{luk} {\L{}UKASIK J. ET AL., {\it Nucl. Instrum. Methods
Phys. Res.}, Sect. A {\bf 587} (2008) 413.}

\bibitem{sfienti} {SFIENTI C. ET AL., {\it Phys. Rev. Lett.} {\bf 102} (2009) 152701.}

\bibitem{traut} {TRAUTMANN W. ET AL., {\it Phys. Rev.} C {\bf 76} (2007) 064606.}

\bibitem{botv} {BOTVINA A. S. ET AL., {\it Nucl. Phys.} {\bf A584} (1995) 737.}


\end{thebibliography}
\end{document}